\documentclass[floatfix,amsmath,aps,prb,eqsecnum]{revtex4-2}
\usepackage{bm}
\usepackage{xcolor}
\usepackage{hyperref}
\usepackage{makeidx}
\usepackage{amsmath}
\usepackage{graphicx}
\usepackage{dcolumn}
\usepackage{amsfonts}
\usepackage{amssymb}
\usepackage[latin9]{inputenc}
\usepackage{color}
\usepackage{subcaption}
\usepackage{float}

\begin{document}

\title{Dehybridization transition in Kondo insulators and heavy fermions}

\author{Mucio A. Continentino}
\email{mucio@cbpf.br}
\affiliation{Centro Brasileiro de Pesquisas F\'{\i}sicas, Rua Dr. Xavier Sigaud 150, 22290-180, Rio de Janeiro, RJ, Brazil}
\date{\today}

\begin{abstract}
In strongly correlated multi-band systems, like inter-metallics, heavy fermions or Kondo insulators, electron-electron and electron-phonon scattering of the electrons in the bands  give rise, at finite temperatures, to a damping of these quasi-particles. This is responsible for producing an effective dehybridization between the electrons in  the large conduction bands and those in the narrow, correlated band. This dehybridization effect has been used to explain the transport properties of inter-metallics  and ARPES experiments in heavy fermions at sufficiently high temperatures. A new insight into this problem has been recently proposed using the theory of non-Hermitian systems. In this note, we review previous work  on dehybridization in Kondo insulators and strongly correlated metals within this new perspective. For this purpose, we  use a parametrization of the self-energy of the strongly correlated electrons obtained from LDA+DMFT calculations. We discuss the nature of the dehybridization transition and its consequences in the electronic spectrum and transport properties.

\end{abstract}

\maketitle

\section{Introduction}

At low temperatures, in  the Fermi liquid regime of heavy fermion systems, the $f$-states of the rare-earth or actinide elements hybridize with the  electrons of the  conduction bands  yielding a coherent  state of heavy quasi-particles. In Kondo insulators this hybridization gives rise to an insulating or semi-metallic ground state. However, as temperature increases many-body effects  and other temperature dependent scattering mechanisms,  as electron-phonon, cause a damping of the electronic quasi-particles and destroy the coherent state. This  occurs due to a dehybridization of the $f$-electronic states and those in the conduction band~\cite{doniach,moriya} caused by  finite life-time effects. In the incoherent high temperature regime the $f$-states behave independently,  decoupled from the conduction band.

The notion that electron-electron or electron-phonon scattering in multi-band systems may alter the hybridization between these bands was discussed in Refs.~\cite{doniach},~\cite{moriya} and~\cite{weger}. At finite temperatures, life-time effects of the quasi-particles make the hybridization less effective producing  eventually a {\it dehybridization-transition}. This approach has been used to understand  transport properties of inter-metallic systems ~\cite{weger} and $f$-bands metals~\cite{freimuth,brodsky,fisk}. Also, the appearance of a gap with decreasing temperature observed in ARPES experiments in the heavy fermion Ce$_2$RhIn$_8$ has been interpreted using this scenario~\cite{ours,cris,note}.

The notion of a quasi-particle in the presence of disorder or interactions is relative. Quite generally, these produce lifetime effects and a general criterion whether this is still a useful concept, compares their energy and damping. In a region where the damping, or inverse lifetime, of these modes is small compared with their energy, they are still well-defined and  the idea of a quasi-particle remains useful. This is the reason why  long wavelength phonons  exist in glasses and spin waves in amorphous ferromagnets~\cite{mucio}. The damping of these excitations due to disorder has a power law dependence on the momentum larger than that of their energy. Then in the hydrodynamic limit $k \rightarrow 0$, they are always well defined. In electronic systems interactions give rise to a damping of the quasi-particles that goes to zero faster than their energy, as the Fermi surface is approached~\cite{mucio}. 

More recently, new insights in the phenomenon  of dehybridization  have been possible due to the progress in the study of non-Hermitian systems and exceptional points~\cite{fu0,fu1,fu2,fu3,fu4,fu5,fu6,fu7}.
In this note we discuss this problem using a semi-phenomenological approach~\cite{weger} that has been validated by microscopic calculations~\cite{moriya,fu4}. We use a parametrization of the self-energy of  the strongly correlated electrons in a  heavy fermion, obtained from LDA+DMFT calculations~\cite{science}, to discuss  the relation between dehybridization and exceptional points of a non-Hermitian Hamiltonian. 

\section{The zero temperature coherent ground state}

At zero temperature the dispersion relations of the quasi-particles of a hybridized two-band  system are given by~\cite{cris}, 
\begin{equation}\label{relhyb}
E_k^{12}=\frac{1}{2} \left[(\epsilon_k^s+\epsilon_k^f) \pm \sqrt{(\epsilon_k^s - \epsilon_k^f)^2 +4|V_k|^2}\right]
\end{equation}
where $\epsilon_k^s$ is the dispersion of the particles in the large conduction band, $\epsilon_k^f$ that in the narrow band  and $V_k$ their hybridization. These quantities should be regarded as   effective parameters, renormalized by interactions in the narrow band, and in this sense incorporate many-body effects. 
\begin{figure}[h]
\centering{\includegraphics[scale=0.5,angle=0]{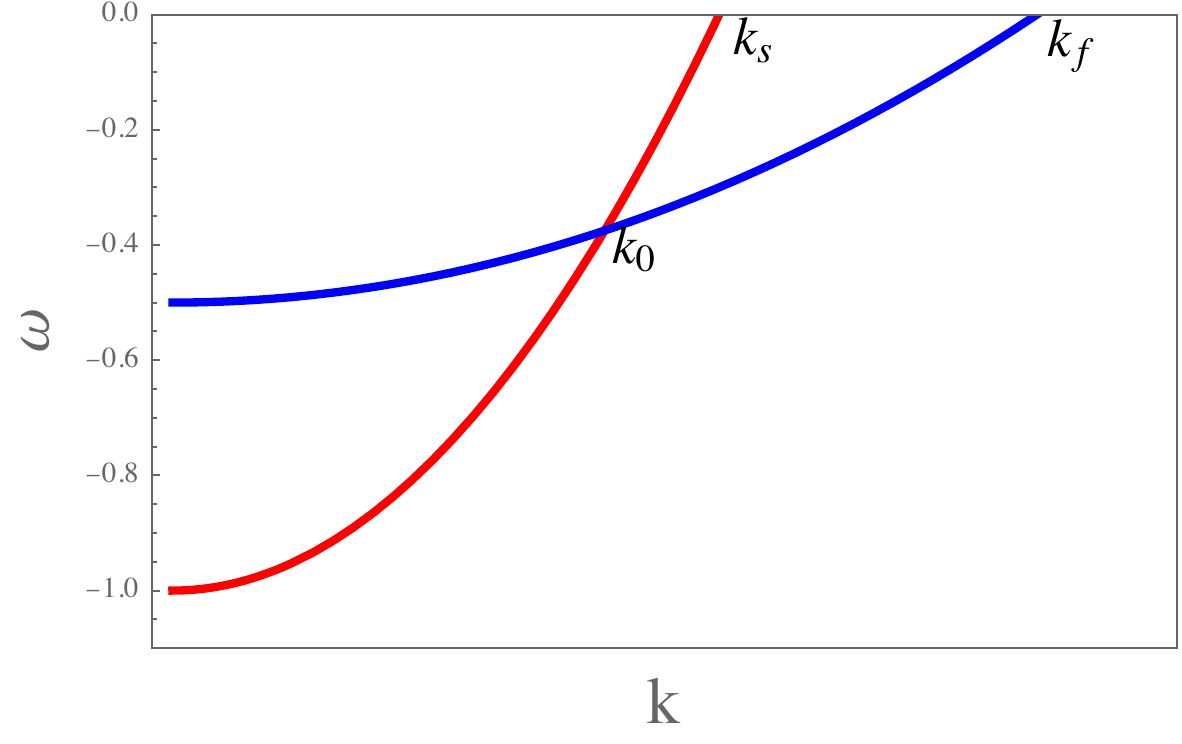}}
\caption{(Color online) Dispersion relations of the system in the absence of hybridization with $\epsilon_f/\mu=0.5$, $\alpha=0.2$ and $\mu=1$.   The Fermi level is at $\omega=0$. We emphasize three special wavevectors, $k_0$ where the bands cross and the Fermi wave-vectors $k_s$ and $k_f$ of the conduction and narrow bands, respectively. } 
\label{fig1}
\end{figure}
The dispersion of the bands, without hybridization are given by,
\begin{eqnarray}
\epsilon_k^s&=&k^2 -\mu \nonumber \\
\epsilon_k^f&=&\alpha k^2 - \epsilon_f,
\end{eqnarray}
where $\alpha< 1$ is the ratio of the effective masses, $\mu$ the chemical potential and $\epsilon_f$ the bottom of the narrow band.
\begin{figure}[h]
\centering{\includegraphics[scale=0.7,angle=0]{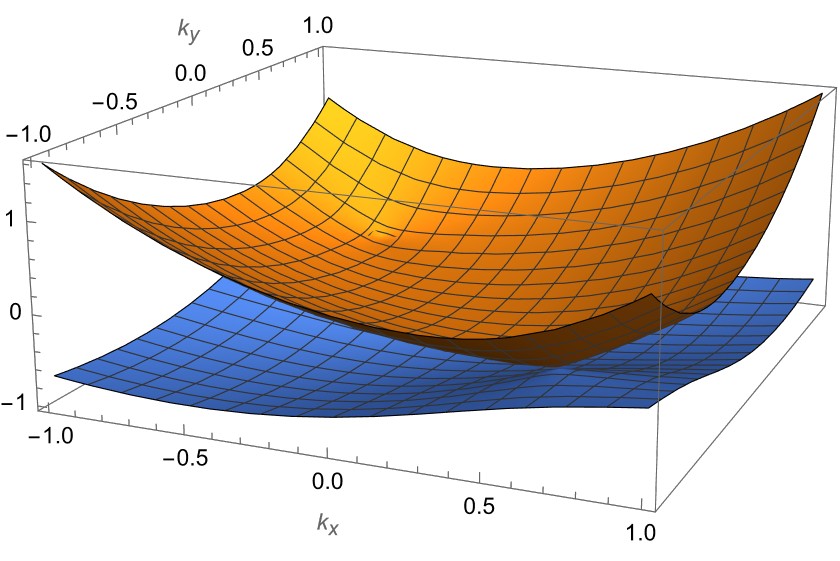}}
\caption{(Color online) Dispersion relation of the hybridized bands, for $v/\mu=0.5$, showing two nodal points along the line $k_y=-k_x$.} \label{fig2}
\end{figure}
The Fermi level is taken at $E_F=0$. In Fig.\ref{fig1} we plot these dispersion relations for $\epsilon_f/\mu=0.5$, $\alpha=0.2$ and $\mu=1$. We emphasize three special wave-vectors in  Fig.~\ref{fig2}.  $k_0$, where the bands cross,  $k_s$ and $k_f$ the Fermi wave-vectors of the  unhybridized $s$ and $f$ bands, respectively.

For the hybridization, we take an anti-symmetric $V_k$, appropriate for the mixing between orbitals of different parities  like $s$-$f$, $p$-$d$ and  $d$-$f$~\cite{anti}. In two dimensions and considering only a small k expansion, we have, $V(\bf{k})$=$iv(k_x+k_y)$. The dispersion relations of the hybridized bands for $v=0.5$ are shown in Fig. \ref{fig2}. All energies are renormalized by $\mu=1$. Notice a pair of  nodal points, at the line $k_y=-k_x$, that appear due to the antisymmetric hybridization.

\section{The dehybridization transition}\label{aha}

In order to fully incorporate the  many-body effects in Eq.~\ref{relhyb}, we have to include the damping of the quasi-particles due to interactions or other scattering mechanisms. Here we consider only that of the heavy quasi-particles due to  interactions. We adopt  the usual  ansatz~\cite{moriya,weger} and introduce an imaginary part in the energy dispersion  of these modes, i.e. $\epsilon_k^f \rightarrow \epsilon_k^f + i \tau_f^{-1}$. This approach has been justified in several papers, including Refs. \cite{moriya} and \cite{fu4}.  We also assume $\tau_f$ is $k$-independent, but has a  temperature dependence.

The energy of the quasi-particles is now given by, 
\begin{equation}
E_k^{12}=\frac{1}{2} \left[(\epsilon_k^s+\epsilon_k^f- i \tau_f^{-1}) \pm \sqrt{(\epsilon_k^s - \epsilon_k^f+ i \tau_f^{-1})^2 +4|V_k|^2}\right]
\end{equation}
with
\begin{equation}
|V_k|^2=v^2 k^2 (1+\sin 2 \theta)
\end{equation}
where we used  $k_x=k \cos \theta$ and $k_y= k \sin \theta$.

\begin{figure}[h]
\centering{\includegraphics[scale=0.35,angle=0]{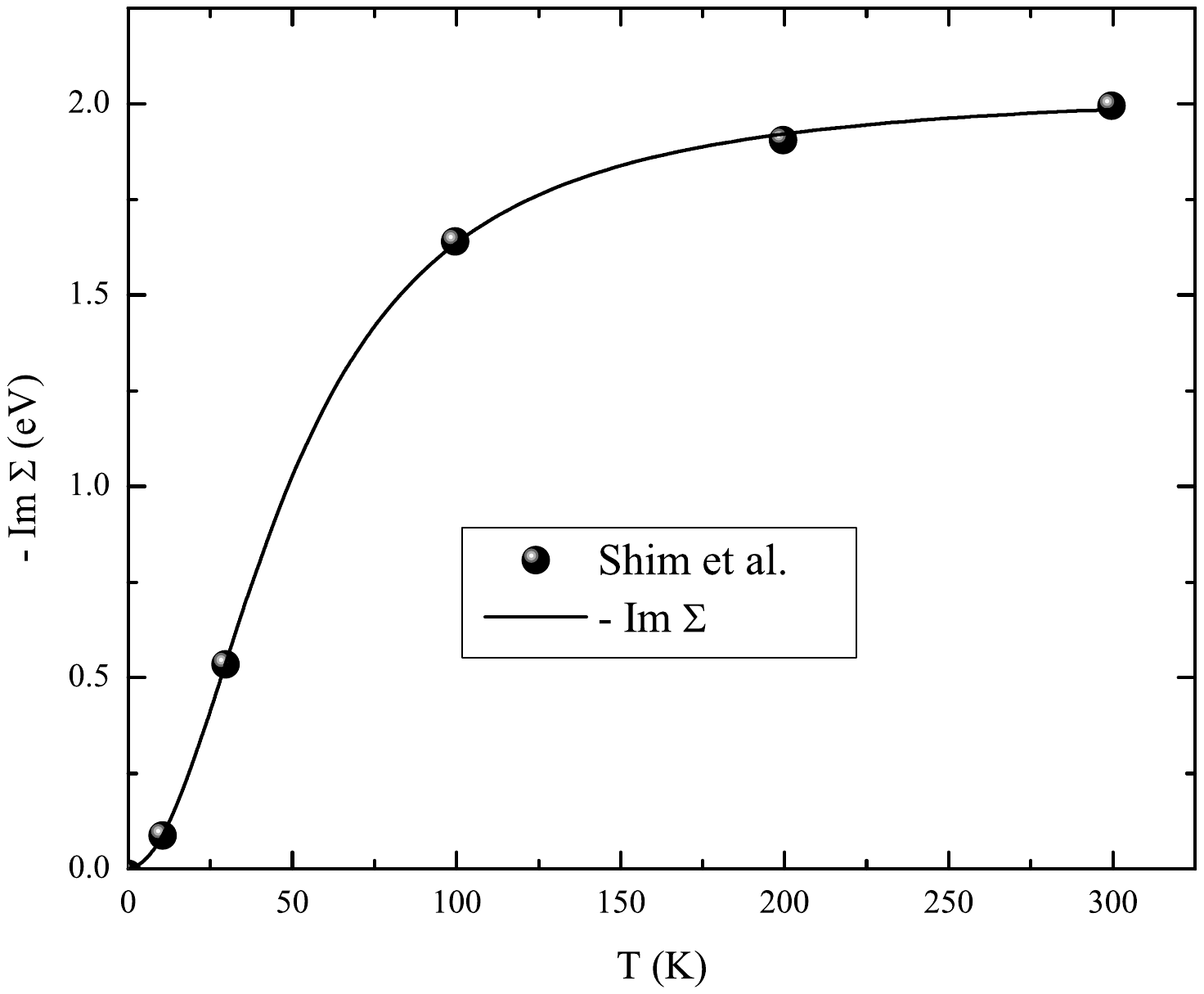}}
\caption{(Color online) The imaginary part of the self-energy of CeIrIn$_5$ as a function of temperature obtained by LDA+DMFT calculations in Ref. \cite{science} (dots). The line is a parametrization of the calculated points given by, $- \Im m\Sigma =a (T/T^*)^2/(1+(T/T^*)^2)$, with $a=2$ eV and $T^*=50$ K. } \label{fig3}
\end{figure}

As pointed out in Refs.~\cite{moriya,fu4}, the lifetime of the $f$-modes due to many-body effects is related  to the imaginary part of their self-energy, 
\begin{equation}
\frac{\hbar}{\tau_f} = -\Im m \Sigma.
\end{equation}
This quantity has been calculated  for the heavy fermion system CeIrIn$_5$  using a LDA+DMFT approach \cite{science}. It is shown in Fig.~\ref{fig3} together with a parametrization of the LDA+DMFT results, given by ($\hbar=1$),
\begin{equation}
\frac{1}{\tau_f} = -\Im m \Sigma = \frac{1}{\tau_{0f}} \frac{(T/T^*)^2}{1+(T/T^*)^2}.
\label{self}
\end{equation}
This function gives an excellent description of the obtained LDA+DMFT results, as can be seen in Fig~\ref{fig3}. It yields, $\hbar \tau_{0f}^{-1}=2$ eV and $T^*=50$K, for this compound. When $T \rightarrow 0$,  the lifetime of the $f$ quasi-particles becomes very large, $\tau_f \rightarrow \infty$,  and the system is in a coherent ground state, with fully hybridized bands. At low temperatures $T \ll T^*$, Eq.~\ref{self} yields the well-known $T^2$ temperature dependence of the inverse lifetime of a  Fermi liquid. The characteristic temperature $T^*$ marks the onset of the Fermi liquid regime.
At high temperatures $T \gg T^*$ the inverse lifetime saturates at a finite value that is essentially the bandwidth $e_f$ of the heavy quasi-particles, $\hbar/\tau_{0f}=\epsilon_f$, as  a consequence of the uncertainty principle~\cite{weger}. 
This saturation is physically important since  at temperatures, for which $\tau_f \sim \tau_{0f}$, the concept of $f$-states as well defined quasiparticles becomes questionable. From the point of view of transport properties~\cite{weger,freimuth}, at these high temperatures, the resistivity due to $s$-$f$ scattering saturates and it begins to be dominated by other scattering mechanisms. The most common is electron-phonon scattering~\cite{weger} that gives rise to a linear temperature dependent resistivity.

We now consider the dehybridization process in more detail for two cases.

\subsection{Nodal semi-metal}

\begin{figure}[h]
\centering{\includegraphics[scale=0.6,angle=0]{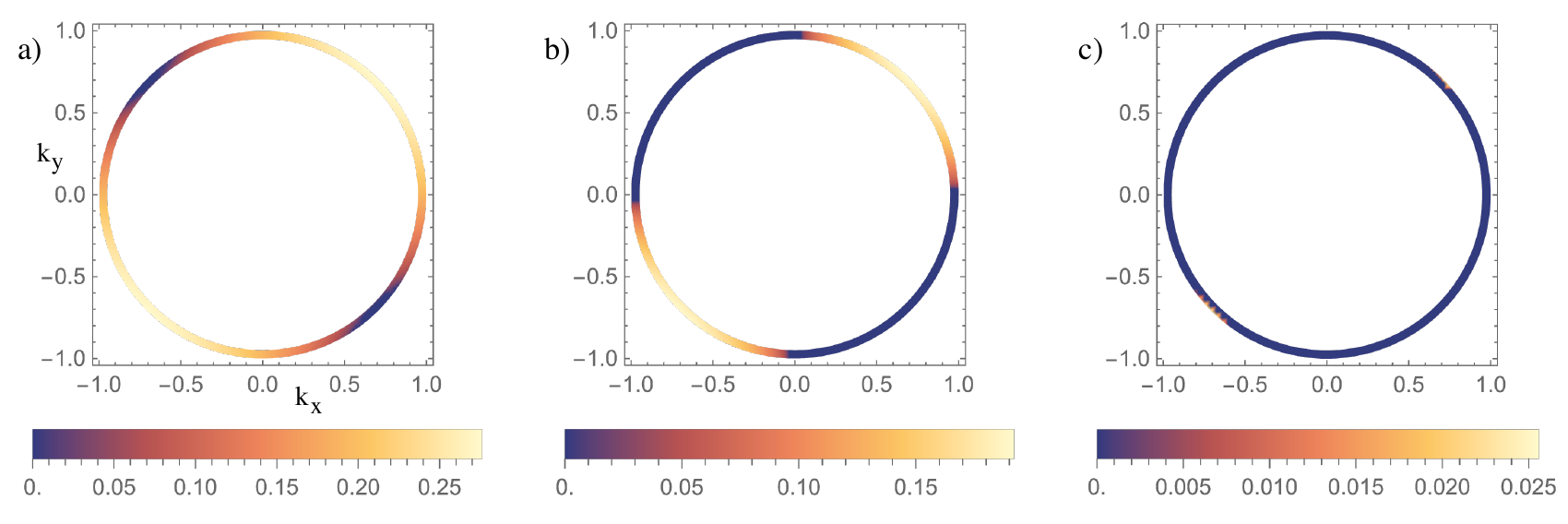}}
\caption{(Color online) Gapless regions in the Fermi {\it surface},  the Fermi arcs,   for different temperatures ($k_0^{2}=\mu$).  a) $T=T_D/8$, b) $T=3T_D/8$ and c) $T=T_D$, with $T_D$ given by Eq. \ref{td}.  } \label{fig4}
\end{figure}

First,  the case where the  bands cross  exactly at the Fermi level, i.e., $\epsilon_{k_0}^s=\epsilon_{k_0}^f=0$ (see Fig.~\ref{fig1}). This implies,  $k_{0}^2=\mu=\epsilon_f/ \alpha$ or $\epsilon_f=\alpha \mu$.
The energy of the quasi-particles is given by,
\begin{equation}
E_{k_0}^{12}=\frac{1}{2} \left[- i \tau_f^{-1} \pm \sqrt{4|V_{k_{0}}|^2 - \tau_f^{-2}} \right]
\end{equation}
where
 $|V_{k_{0}}|^2 =v^2 k_0^2(1+\sin2 \theta)$, with $0 \le |V_{k_0}|^2 \le 2v^2 k_0^2$.
At $T=0$ with $\tau_f=\infty$, the coherent ground state  has two nodes at the Fermi {\it surface} for $\theta=-\pi/4$, i.e along the line $k_y=-k_x$, as shown in Fig.\ref{fig2}. The ground state is a nodal semi-metal. As temperature increases, $\tau_f^{-1}$ increases and the Fermi points develop into Fermi arcs, as shown in Fig.~\ref{fig4}. We used the parametrization for $\tau_f^{-1}$ given by Eq.~\ref{self}.

As pointed out in Ref.~\cite{fu4} dehybridization in this case occurs gradually, with a different dehybridization temperature for each angle $\theta$ in the Fermi surface. As temperature increases, the Fermi arcs increase until the whole Fermi surface becomes gapless at a  {\it dehybridization temperature} $T_D$, when  $\sin 2 \theta=1$, as shown in  Fig.~\ref{fig4}c.  
The appearance of the Fermi arcs is connected with the existence of exceptional points  where the energies $E^1(k)=E^2(k)$,  become degenerate~\cite{fu4}. They are also a direct consequence of the angular or momentum dependence of the hybridization, which determines the points of the Brillouin zone where they appear.  
Dehybridization in this case is a gradual process, a crossover phenomenon, whose energy scale  can be characterized, for example,  by the {\it dehybridization temperature} $T_D$ at which, the whole Fermi surface becomes gapless. This  occurs   for $ \tau_{f}(T_D)^{-2}=8v^2 k_0^2$. and using $k_0^2=\mu$, $\tau_{0f}^{-1}=\epsilon_f$, we obtain,
\begin{equation}\label{td}
T_D=T^* \sqrt{\frac{v^*}{1-v^*}},
\end{equation}
with $v^*=2\sqrt{2}(v/\epsilon_f)$.
For $v/\epsilon_f=0.1$, we find $T_D \approx 1.14  T^*$.
Notice that dehybridization  occurs at  exceptional points of the non-Hermitian Hamiltonian describing the system~\cite{fu4}. As shown in Eq.~\ref{td},  the crossover  temperature $T_D$  is  proportional to $T^*$,  an important point that we discuss in more detail below.

\subsection{Metal}

We now consider the case where the ground state is metallic. This can be described assuming that the line the  dispersion relations cross is not the Fermi line.  We  take,  $\epsilon_{\tilde{k}_0}^s=\epsilon_{\tilde{k}_0}^f=\epsilon=(\alpha \mu-\epsilon_f)/(1- \alpha)$, at the crossing line  given by $\tilde{k}_0^2=(\mu-\epsilon_f)/(1-\alpha)$. 
The energies of the quasi-particles of the hybridized system are given by,
\begin{equation}
E_k^{12}=\frac{1}{2} \left[2 \epsilon- i \tau_f^{-1} \pm \sqrt{4|V_{\tilde{k}_{0}}|^2 - \tau_f^{-2}} \right]
\end{equation}
where, $|V_{\tilde{k}_{0}}|^2 =v^2 \tilde{k}_0^2(1+\sin2 \theta)$.
\begin{figure}[h]
\centering{\includegraphics[scale=0.5,angle=0]{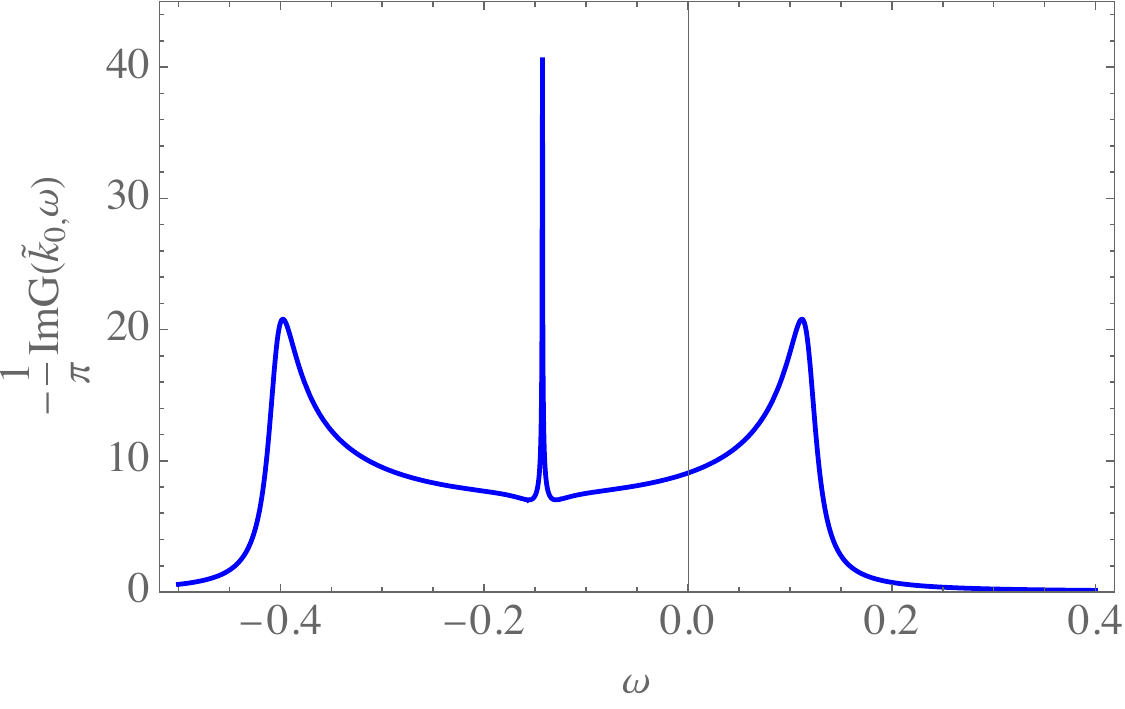}}
\caption{(Color online) The average spectral at $\tilde{k}_0$ for $T=0.2T^{*}$. } \label{fig5}
\end{figure}
\begin{figure}[h]
\centering{\includegraphics[scale=0.5,angle=0]{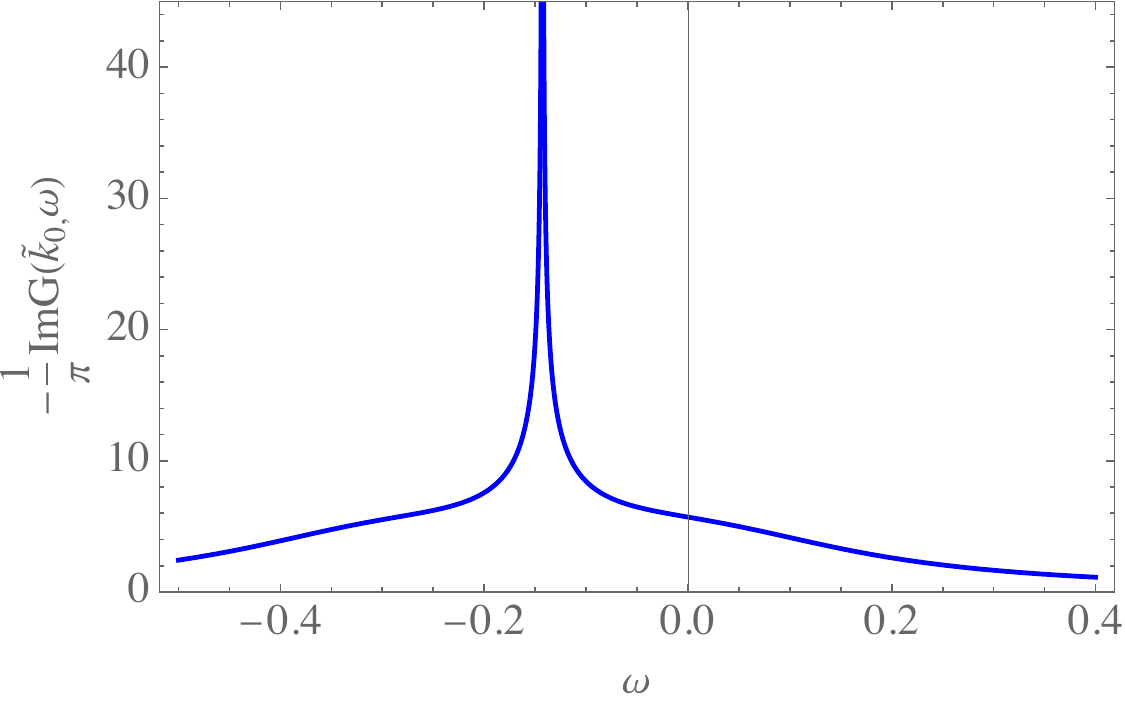}}
\caption{(Color online) The average spectral at $\tilde{k}_0$ for $T=2 T^{*}$. } \label{fig6}
\end{figure}
 
The frequency dependent spectral density at the wavevector $\tilde{k}_0$ where the bands cross is given by the imaginary part of the Green's functions~\cite{mucio},  
\begin{equation}
-\frac{1}{\pi} \Im m G(\tilde{k}_0, \omega)= -\frac{1}{\pi} \sum_{i=1,2}\Im m ( \frac{1}{\omega - E_{\tilde{k}_0}^i}).
\end{equation}
The average of this quantity over the Fermi surface is shown in Figs.~\ref{fig5} and~\ref{fig6}  for low, $T=0.2 T^*$, and high temperatures, $T=2 T^*$,  respectively. The parameters we use are, $\mu=1$, $\alpha=0.3$, $v=0.2$ and $\epsilon_f=0.4$. For low temperatures, $T=0.2 T^*$, the spectral density  at $\tilde{k}_0$, has a peak close to the energy $\epsilon$ and two peaks separated by repulsion due to hybridization.  As temperature  increases,  dehybridization occurs.  The contribution of the $f$-states has a tendency to accumulate close to $\epsilon$ while the conduction states remain spread in energy, as shown in Fig.~\ref{fig6} for $T=2 T^*$. When $\alpha \rightarrow 0$, i.e., for a nearly localized $f$-level, $\epsilon= (\alpha \mu-\epsilon_f)/(1- \alpha) \approx - \epsilon_f$. In this case the spectral density at $\tilde{k}_0$, which contributes to the density of states, can become very large as $\epsilon$ gets closer to the Fermi level. This can be related to a tendency of the system towards a magnetic instability. 
The crossover between the  two regimes is set by the characteristic temperature $T_D \sim T^*$, at which 
significant changes in the spectral weight occur.

\section{Discussion}

What is  the nature of the dehybridization transition discussed here?  Is it a temperature driven phase transition or a crossover phenomenon? When the momentum dependence of the hybridization is taken into account,  dehybridization occurs gradually and it is clearly a crossover from  mixed to pure states~\cite{fu4}. This crossover manifests in  the presence of Fermi arcs that increase with temperature  to a point where eventually the whole Fermi surface is gapless, or dehybridized. 
Even in the case of a $k$-independent hybridization, due to the finite lifetime of the heavy quasiparticles, there is a finite temperature range for dehybridization to occur. Then, there is no evidence for a sharp dehybridization phase transition, which  instead should be identified as a crossover phenomenon. 
\begin{figure}[h]
\centering{\includegraphics[scale=1.2,angle=0]{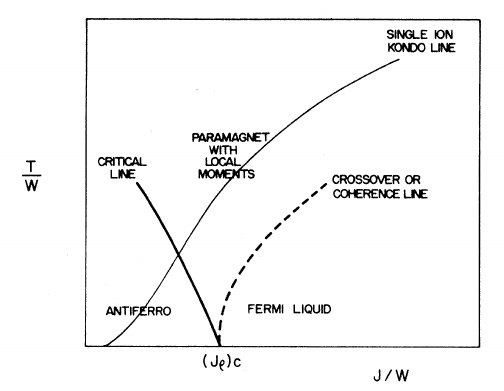}}
\caption{ Phase diagram of heavy fermion systems as described by the Kondo lattice model~\cite{coherence}. $J$ is the Kondo interaction between local and conduction electrons, $W=1/\rho$ is the bandwidth of the conduction band.  The quantum critical point, that separates the long range ordered magnetic state from the Fermi liquid ground state is at $J/W=(J\rho)_c$. Reprinted from Mucio A. Continentino, Gloria M. Japiassu and Am\'os Troper, 1989, Critical approach to the coherence transition in Kondo lattices, Phys. Rev. B39: 9734-9737.} \label{fig7}
\end{figure}

Another relevant point concerns whether the dehybridization crossover temperature  is a new energy scale of the  correlated multi-band many-body system.
For the well studied  heavy fermion compounds of the families Ce$_m$MIn$_{3m+2}$ (M = Co, Rh, Ir; and
m = 1,2;)  \cite{ours,cris}, the characteristic temperature $T^*$ and the dehybridization temperature $T_D$ are much higher than the critical temperatures of the  magnetic or superconducting phase transitions.
In particular the $T^*$ obtained in Section \ref{aha} for CeIrIn$_5$ is much higher than the superconducting critical temperature of this system~\cite{science,pagliuso}. 
Besides,  in the LDA+DMFT calculations, the characteristic temperature, identified in this case with a crossover from the localized to the itinerant state~\cite{science},   arises from a parametrization of the momentum independent imaginary part of the  self energy of the $f$-electrons, which captures local correlations between quasi-particles. Then, dehybridization is a relative high temperature phenomenon compared to any ordering temperature in these systems. It occurs in an incoherent regime where correlations between the different types of quasi-particles are essentially local. In this case it seems natural to identify the characteristic temperature with the Kondo temperature  of the material, i.e. $T^*=T_K$, since the latter marks the onset of the Fermi liquid regime.  This is corroborated by the results of Ref.~\cite{fu2} showing that exceptional points, associated with the dehybridization crossover appear at the Kondo temperature, at which  magnetic moments  are screened.

A different situation occurs if the ground state of the system is a coherent hybridized Fermi liquid  close to the quantum critical point (QCP) of a magnetic instability, like in many heavy fermion materials~\cite{coherence,livro}(see Fig~\ref{fig7}). The relevant scale  for the onset of the Fermi liquid regime  in the non-critical side of the phase diagram is the coherence temperature~\cite{coherence}, $T_{coh}=|g|^{\nu z}$, as shown in Fig~\ref{fig7}. Here $g$ is the distance to the QCP and $\nu$ and $z$ the correlation length and dynamical critical exponents, respectively~\cite{coherence}.  Notice in Fig~\ref{fig7} that, although in this region of the phase diagram there is a higher energy scale given by the Kondo temperature, in this case it is the coherence temperature that gives the scale for dehybridization, since it marks the crossover to the Fermi liquid regime.

Eq.~\ref{self} for the temperature dependence of the imaginary part of the self energy turned out to be a very good description of the LDA+DMFT results for CeIrIn$_5$. Its  main feature  is to show that this quantity  scales with temperature as a function $f(T/T^*)$, and has the correct asymptotic behaviors, $f(T/T^* \rightarrow 0) \sim (T/T^*)^2$ and $f(T/T^* \rightarrow \infty) \sim (1/\tau_{0f})$, a constant. The temperature $T^*$ marks the onset of the Fermi regime and can be identified either with the Kondo or the coherence temperature depending on the region of the phase diagram. In any case,   the dehybridization temperature $T_D$  is related to one of these temperatures and is not a new energy scale of the system.

\section{Conclusions}

In this note we reviewed and presented results for the phenomenon of thermal induced dehybridization in multi-band correlated electronic systems. This approach gives a good description of the finite temperature transport and spectroscopic properties of these materials. Besides, it is consistent with the results of numerical treatments as the LDA+DMFT method. A new perspective of this problem is provided by  the theory of non-Hermitian systems that is throwing light on the physics of this phenomenon.

We used a phenomenological approach substantiated by analytical results~\cite{moriya}, LDA-DMFT~\cite{science} and DMFT~\cite{fu4} calculations.
The dehybridization transition is not a temperature driven phase  transition, even in the extreme case of a k-independent hybridization. It is  a crossover phenomenon related to gradual modifications of the Fermi surface and in metals of rearrangements of the spectral weight. 
We  argued  that the dehybridization temperature  is not a  new energy scale in the problem of mixed correlated materials. It can be identified with the Kondo  or coherence temperature depending on the region of the phase diagram. In the present approach we did not consider the possibility of a zero temperature dehybridization transition. 

\section{Acknowledgements}
I would like to thank the Brazilian National Council for Scientific and Technological Development (CNPq),  Grant Number 305810/2020-0, and the Foundation for Support of Research in the State of Rio de Janeiro (FAPERJ), Grant Number 201223/2021, for partial financial support.

%
%
%
%


\begin{thebibliography}{99}


\bibitem{doniach} S. Doniach, Many Electron Effects in the Actinides, in {\it The Actinides}, ed. A. J. Freeman and
J. B. Darby (Academic Press, New York, 1974)
Vol.II, p. 51.

\bibitem{moriya} Y.Takaoka and T. Moriya, Theoretical Study of Electrical Resistivity in Actinide Metals, J. Phys. Soc. Japan, {\bf 52}, 605 (1983).

\bibitem{weger} M. Weger, Dehybridization transition in
intermetallic transition-metal
compounds, Phil. Mag. \textbf{B52}, 701 (1985).

\bibitem{freimuth} A. Freimuth, Correlation between transport properties and quasielastic
linewidths of Ce and Yb compounds with unstable 4f-shell, JMMM, {\bf 68}, 28 (1987).

\bibitem{brodsky} A. J. Arko, F. Y. Fradin, and M. B. Brodsky, Magnetic, Transport, and Nuclear-Magnetic-Resonance Properties of U$_{1-x}$Pu$_x$Al$_2$, Phys.Rev. {\bf B8}, 4104 (1973); E. V. Sampathkumaran and R. Vijayaraghavan, Evidence for 4f-Ligand Dehybridization in the Evolution of Heavy-Fermion
Behavior in the Series CeCu$_{2-x}$Ni$_x$Si$_2$, Phys. Rev. Lett. {\bf 56}, 2861 (1986).


\bibitem{fisk} D. Leuenberger, J. A. Sobota, S.-L. Yang, H. Pfau, D. J. Kim, S.-K. Mo, Z. Fisk, 
P. S. Kirchmann, and Z.-X. Shen, Dehybridization of f and d states in the heavy-fermion system YbRh$_2$Si$_2$, Phys. Rev. {\bf B97}, 165108 (2018).

\bibitem{ours} C. Adriano, F. Rodolakis, P. F. S. Rosa, F. Restrepo, M. A.
Continentino, Z. Fisk, J. C. Campuzano, and P. G. Pagliuso, Unveiling the hybridization gap in Ce2RhIn8 heavy fermion compound, arXiv:1502.02544 [cond-mat.str-el], https://doi.org/10.48550/arXiv.1502.02544.

\bibitem{cris} C. Adriano, C. Giles, E. M. Bittar, L. N. Coelho, F. de Bergevin, C. Mazzoli, L. Paolasini, W. Ratcliff, R. Bindel, J. W. Lynn, Z. Fisk, and P. G. Pagliuso, Cd doping effects in the heavy-fermion compounds Ce$_2$MIn$_8$ (M=Rh and Ir),  Phys. Rev. \textbf{B81}, 245115 (2010); M. Nicklas, V. A. Sidorov, H. A. Borges, P. G. Pagliuso, C. Petrovic, Z. Fisk, J. L. Sarrao, and J. D. Thompson, Magnetism and superconductivity in Ce$_2$RhIn$_8$, Phys.  Rev.  {\bf B 67}, 020506 R (2003).


\bibitem{note} The above references do not exhaust the works on effects of dehybridization in metallic system.

\bibitem{mucio} {\it Key Methods and Concepts in Condensed Matter Physics: Green's functions and real space
renormalization group}, IOP Expanding Physics, IOP Publishing, Bristol, UK,  2021.


\bibitem{fu0}V. Kozii and L. Fu, Non-Hermitian topological theory of
finite-lifetime quasiparticles: Prediction of bulk Fermi arc
due to exceptional point, arXiv:1708.05841.


\bibitem{fu1} H. Shen, B. Zhen, and L. Fu, Topological Band Theory for
Non-Hermitian Hamiltonians, Phys. Rev. Lett. 120, 146402
(2018).

\bibitem{fu2} T. Yoshida, R. Peters, and N. Kawakami, Non-Hermitian
perspective of the band structure in heavy-fermion systems,
Phys. Rev. B 98, 035141 (2018);Yoshihiro Michishita, Tsuneya Yoshida, and Robert Peters, Relationship between exceptional points and the Kondo effect in f -electron materials, Phys. Rev.  {\bf B 101}, 085122 (2020).

\bibitem{fu3} M. Papaj, H. Isobe, and L. Fu, Nodal arc of disordered Dirac
fermions and non-Hermitian band theory, Phys. Rev. B 99,
201107(R) (2019).

\bibitem{fu4} Yuki Nagai,  Yang Qi,  Hiroki Isobe, Vladyslav Kozii,  and Liang Fu, DMFT Reveals the Non-Hermitian Topology and Fermi Arcs in Heavy-Fermion Systems, Phys. Rev. Lett, {\bf 125}, 227204 (2020).

\bibitem{fu5} Y. Michishita, T. Yoshida, and R. Peters, Relationship
between exceptional points and the Kondo effect
in f-electron materials, Phys. Rev. B 101, 085122
(2020).

\bibitem{fu6} Haoyu Hu,  Lei Chen, Jian-Xin Zhu, Rong Yu, and Qimiao Si, Orbital-selective Mott phase as a dehybridization fixed point, arXiv:2203.06140 [cond-mat.str-el].


\bibitem{fu7} Vladyslav Kozii and Liang Fu, Non-Hermitian topological theory of finite-lifetime quasiparticles: Prediction
of bulk Fermi arc due to exceptional point, Phys. Rev.  {\bf B109}, 235139 (2024).

\bibitem{anti} Mucio A. Continentino, Fernanda Deus ,Igor T. Padilha and Heron Caldas, Topological transitions in multi-band superconductors, Annals of Physics {\bf 348}, 1  (2014).

\bibitem{science}  J. H. Shim, K. Haule, G. Kotliar, Science \textbf{318}, 1615 (2007).

\bibitem{pagliuso} C. Petrovic, R. Movshovich, M. Jaime, P. G. Pagliuso, M. F. Hundley, J. L. Sarrao, Z. Fisk and J. D. Thompson, A new heavy-fermion superconductor CeIrIn5: A relative of the cuprates?,  Europhys. Lett. 53, 354 (2001).

\bibitem{coleman} P Coleman, Mixed valence as an almost broken symmetry,
Phys. Rev. {\bf B35},  5072 (1987); P Coleman and N Andrei, Kondo-stabilised spin liquids and heavy fermion superconductivity, Journal of Physics: Condensed Matter, {\bf 26}, 4057 (1989).


\bibitem{coherence} Mucio A. Continentino, Gloria M. Japiassu and Am\'os Troper, Critical approach to the coherence transition in Kondo lattices, Phys. Rev. {\bf B39}, 9734 (1989).

\bibitem{livro} {\it Quantum scaling in many-body systems: an approach to quantum phase transitions}, Mucio A. Continentino, Cambridge University Press, UK, 2017.

\end{thebibliography}
\end{document}